\documentclass{article}

\usepackage{microtype}
\usepackage{graphicx}
\usepackage{subcaption}
\usepackage{booktabs} 

\usepackage{hyperref}

\usepackage[preprint]{icml2026}

\usepackage{amsmath}
\usepackage{amssymb}
\usepackage{mathtools}
\usepackage{amsthm}

\usepackage{listings}
\usepackage[capitalize,noabbrev]{cleveref}
\usepackage{enumitem}

\theoremstyle{plain}

\theoremstyle{definition}

\theoremstyle{remark}

\usepackage[textsize=tiny]{todonotes}

\icmltitlerunning{GetBatch: Distributed Multi-Object Retrieval for ML Data Loading}

\begin{document}

\twocolumn[
  \icmltitle{GetBatch: Distributed Multi-Object Retrieval for ML Data Loading}

  \icmlsetsymbol{equal}{*}

  \begin{icmlauthorlist}
    \icmlauthor{Alex Aizman}{comp}
    \icmlauthor{Abhishek Gaikwad}{comp}
    \icmlauthor{Piotr Żelasko}{comp}
  \end{icmlauthorlist}

  \icmlaffiliation{comp}{NVIDIA, Santa Clara, USA}

  \icmlcorrespondingauthor{Alex Aizman}{aaizman@nvidia.com}
  \icmlcorrespondingauthor{Abhishek Gaikwad}{abhgaikwad@nvidia.com}
  \icmlcorrespondingauthor{Piotr Żelasko}{pzelasko@nvidia.com}

  \icmlkeywords{Machine Learning, ICML}

  \vskip 0.3in
]

\printAffiliationsAndNotice{}  

\begin{abstract}

Machine learning training pipelines consume data in batches. 
A single training step may require thousands of samples drawn from shards distributed across a storage cluster. 
Issuing thousands of individual GET requests incurs per-request overhead that often dominates data transfer time.
To solve this problem, we introduce GetBatch - a new object store API that elevates batch retrieval to a first-class storage operation, replacing independent GET operations with a single deterministic, fault-tolerant streaming execution.
GetBatch achieves up to 15$\times$ throughput improvement for small objects and, in a production training workload, reduces P95 batch retrieval latency by 2$\times$ and P99 per-object tail latency by 3.7$\times$ compared to individual GET requests.

\end{abstract}

\section{Introduction}
\label{sec:intro}

The growing scale of data and training workflows forced the ML community to move away from in-memory data. 
Modern workflows are characterized by a data loader abstraction that feeds batches of data into a model's training loop.
However, depending on the scale, different data loading mechanisms must be employed.

The two main approaches are based on either random access or sequential data access patterns.
With "small enough" data, a random-access data loader also known as a "map-style dataset," is desirable--and typically seen in most data loading examples in frameworks such as PyTorch.
The benefit of the random access approach is that it is easy to maintain almost perfect sampling randomness, at some efficiency cost.
At a "large enough" data scale, that efficiency cost becomes a major bottleneck in model training: when data size grows beyond local storage, object stores must be adopted.
However, object stores impose a high overhead for establishing a GET request per retrieved sample.
Therefore, to reduce that overhead, scalable data loading typically employs sequential I/O by packaging data samples into groups called shards---for example, using TAR archives. 
However, sequential I/O makes it more difficult to preserve the same high level of sampling randomness as random access I/O, requiring additional complexity such as shard order shuffling, data blending with stream multiplexers, or in-memory shuffling buffers.

\begin{figure*}
    \centering
    \begin{subfigure}[b]{0.48\textwidth}
        \centering
        \includegraphics[width=\linewidth]{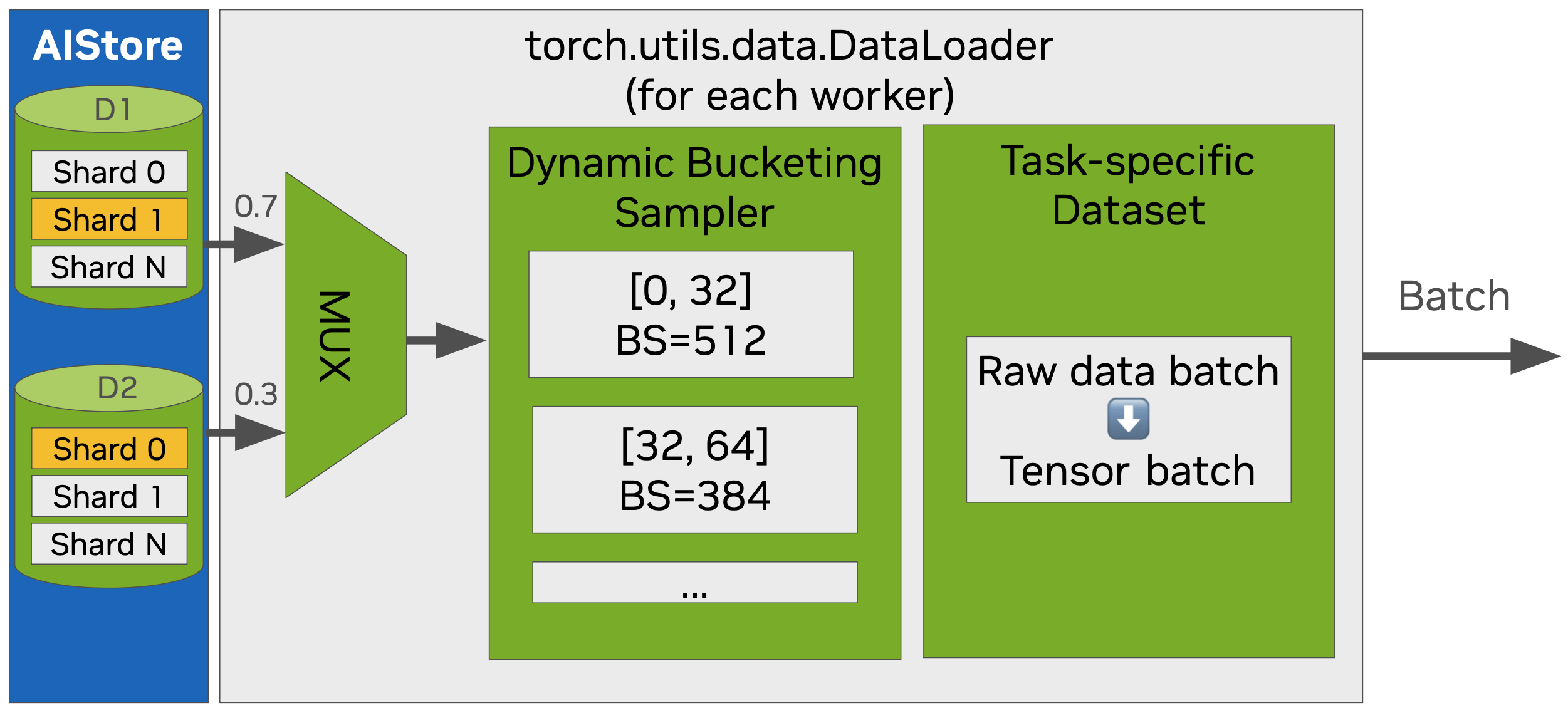}
        \caption{Sequential I/O data loading flavor. The data sampler selects a shard to read, and it must be read entirely, left-to-right. To improve the randomness, the sampler interleaves multiple shards, and prefills examples into a buffer. The sampling is further stratified on factors such as sequence length via dynamic bucketing.}
        \label{fig:aistore_sequential_io}
    \end{subfigure}
    \hfill
    \begin{subfigure}[b]{0.48\textwidth}
        \centering
        \includegraphics[width=\linewidth]{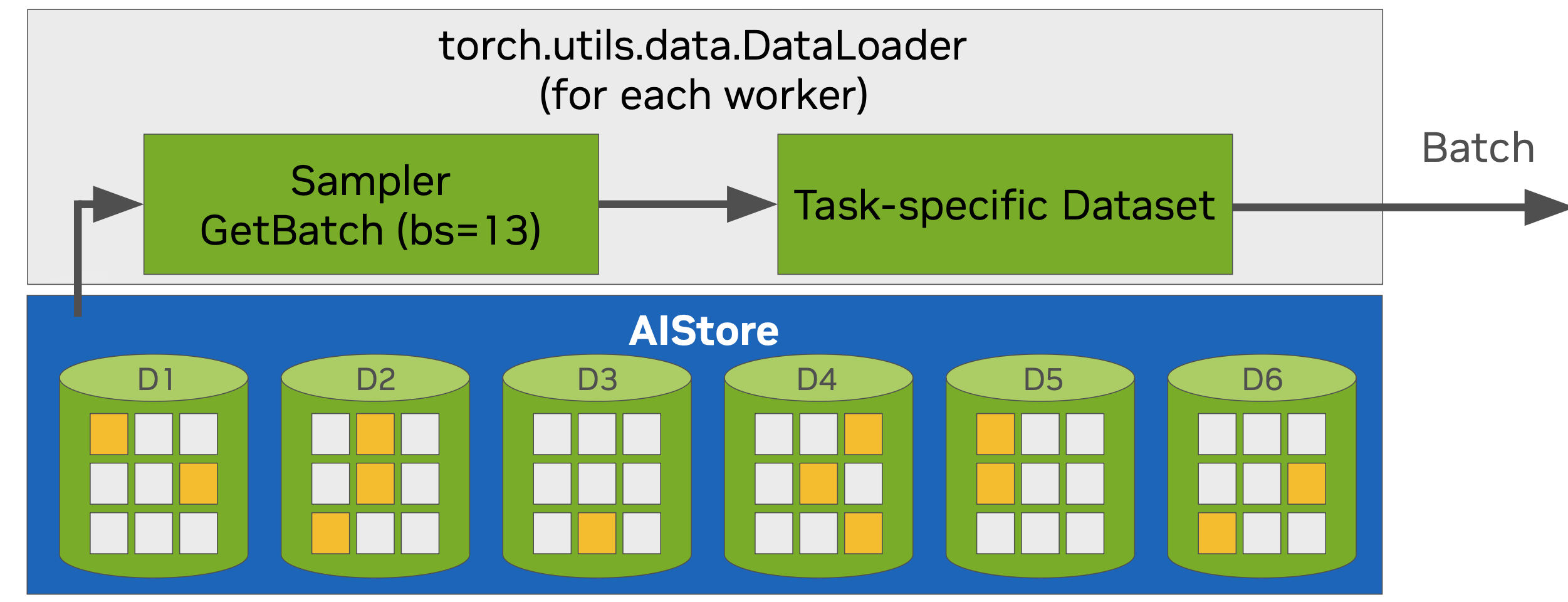}
        \caption{Random access I/O data loading flavor. The data sampler is able to select any example in the entire dataset, regardless of its physical layout, and optionally stratify the sampling prior to data access. Normally, each object must be fetched through a separate request from the data loading client with substantial overhead. GetBatch efficiently batches data access into a single request.}
        \label{fig:aistore_random_io}
    \end{subfigure}
    \caption{AIStore sequential and batched random access data loading patterns. Sequential I/O (a) reads entire shards and selects samples from a buffer, while GetBatch (b) retrieves only the sampled items in a single request.}
    \label{fig:aistore_comparison}
\end{figure*}

We propose an alternative approach that brings the best of both worlds: a batched random-access I/O mechanism called GetBatch.
The core idea is to enable the client code to sample an arbitrary batch, and then retrieve its corresponding data through a single GetBatch request sent to the object store. 
The object store internally fetches each example concurrently, possibly from multiple shards, disks, and machines, and streams it back to the client as a single tar archive.
The differences between sequential I/O and GetBatch are illustrated in \Cref{fig:aistore_comparison}.
We implement GetBatch in NVIDIA AIStore object store and observe up to 15x data loading speedup compared to random access I/O across synthetic benchmarks and a production-scale training workload. Conceptually, GetBatch elevates batch retrieval to a first-class storage primitive by replacing thousands of independent object reads with a single deterministic, fault-tolerant streaming operation that preserves request order in the output stream.

Our main contributions are:
\begin{enumerate}[noitemsep, topsep=4pt]
    \item A \textbf{storage-native batched retrieval primitive} for ML data loading that combines the sampling flexibility of random access with the throughput of sequential I/O.
    \item A \textbf{distributed execution model} with Designated Target (DT) coordination that parallelizes retrieval across cluster nodes while preserving strict output ordering.
    \item \textbf{Empirical evaluation} demonstrating up to 15$\times$ throughput improvement on synthetic benchmarks and stable throughput in production-scale training.
    \item An \textbf{open-source implementation} in NVIDIA AIStore with Python SDK integration for existing training frameworks.
\end{enumerate}

\section{Methods}
\label{sec:methods}
This section describes the architecture and execution model underlying GetBatch. We begin with an overview of AIStore (\Cref{sec:methods:aistore}), then present the design and execution semantics of GetBatch (\Cref{sec:getbatch:design}), its distributed server-side execution (\Cref{sec:getbatch:server}), execution options (\Cref{sec:getbatch:options}), and the client-side interface (\Cref{sec:methods:getbatch_client_side}).

\subsection{AIStore}
\label{sec:methods:aistore}

AIStore (AIS) is a lightweight, high-performance distributed object storage system tailored for large-scale AI workloads. Built for linear scale-out and balanced I/O distribution, AIStore can be deployed on anything from a single Linux machine to a bare-metal cluster of arbitrary size. It supports multi-cloud access, native bucket and object semantics, erasure coding, and n-way mirroring for reliability, along with a comprehensive HTTP API compatible with Amazon S3 clients~\cite{aistore_github}.

AIStore provides advanced capabilities for ML pipelines including archive operations (TAR, ZIP), inline and offline data transformations, and the batching API for multi-object retrieval described in this paper. This paper focuses on the execution semantics and distributed coordination required to support batched retrieval at scale; API syntax and operational usage are documented in AIStore's public documentation~\cite{aistore_github}.

\subsection{GetBatch Design and Execution Semantics}
\label{sec:getbatch:design}

Machine learning training pipelines consume data in \emph{batches}: a single training step may require hundreds or thousands of samples, often drawn from archive shards distributed across a storage cluster. Conventional distributed storage systems expose retrieval at file or object granularity, forcing clients to issue thousands of independent read (or GET) requests per batch. The resulting per-request overhead—network round trips, per-request control-plane processing, and connection management—often dominates data transfer time and leaves compute resources underutilized.

GetBatch addresses this mismatch by elevating batch retrieval to a first-class storage primitive. Instead of issuing many independent requests, a client submits a single request specifying the needed data items; the storage system assembles them across the cluster and delivers the result as a single serialized output stream (default: uncompressed TAR archives). One request, one response, with no designed-in limit on the number of data items.

Implementation-wise, a GetBatch request is issued as an HTTP GET with a JSON body that specifies the output format, retrieval entries, and execution options. Although request bodies in GET operations are uncommon, they are permitted by HTTP semantics~\cite{rfc9110} and are required here to encode large request lists that exceed practical URL length limits.

A single GetBatch request may span multiple buckets and object namespaces, allowing training pipelines to assemble composite samples (e.g., features, labels, metadata) without issuing separate requests or performing client-side joins.
{\footnotesize
\begin{verbatim}
{
  "mime": "tar",
  "in": [
    {"bucket": "imagenet",
     "objname": "images/img_0001.jpg"},
    {"bucket": "imagenet",
     "objname": "images/img_0002.jpg"},
    {"bucket": "shards",
     "objname": "train-0003.tar",
     "archpath": "labels/0003.txt"},
    {"bucket": "shards",
     "objname": "train-0003.tar",
     "archpath": "images/0003.jpg"}
  ],
  "strm": true,
  "coer": false,
  "coloc": 2
}
\end{verbatim}
}

GetBatch enforces strict output ordering: response entries appear in the exact same order as the request list, regardless of whether individual items originate from standalone objects or archive shards, or where they are physically located in the cluster. When continue-on-error mode is enabled and an entry cannot be retrieved, the output stream contains an explicit placeholder preserving positional correspondence (see \Cref{sec:getbatch:options}). This ordering guarantee enables deterministic alignment between training samples and labels without client-side reassembly and is essential for reproducible training.

The client's request arrives at an arbitrary AIS gateway (typically, via standard load balancing). The gateway selects a storage node to serve as the Designated Target (DT) for this request—either randomly or, when colocation hints are provided, by choosing the node that owns the most requested items (see \Cref{sec:getbatch:options}). The DT then assumes responsibility for assembling the serialized output stream, maintaining per-request execution state and enforcing output order, while other storage nodes act as senders, reading locally owned items and delivering them to the DT. This design distributes storage I/O and network transfers across the cluster while centralizing only the final serialization step.

\subsection{GetBatch Server-side Execution}
\label{sec:getbatch:server}

A GetBatch request is executed as a distributed operation coordinated by a single Designated Target (DT). Retrieval of individual data items is parallelized across storage nodes, while ordering and final assembly are handled exclusively by the DT.

\subsubsection{Execution Flow}

Clients issue GetBatch requests to any proxy, a stateless gateway node that exposes the storage API. Upon receiving a request, the proxy selects a DT using consistent hashing under the current cluster membership. This default routing strategy assumes batch entries are uniformly distributed and minimizes coordination overhead; when explicit colocation hints are provided, the proxy may apply an alternative DT selection strategy to reduce cross-node transfers.

Execution proceeds in three phases: (1) DT registration, (2) distributed sender activation, and (3) client redirection and ordered assembly.

\begin{figure}[t]
  \centering
  \includegraphics[width=\linewidth]{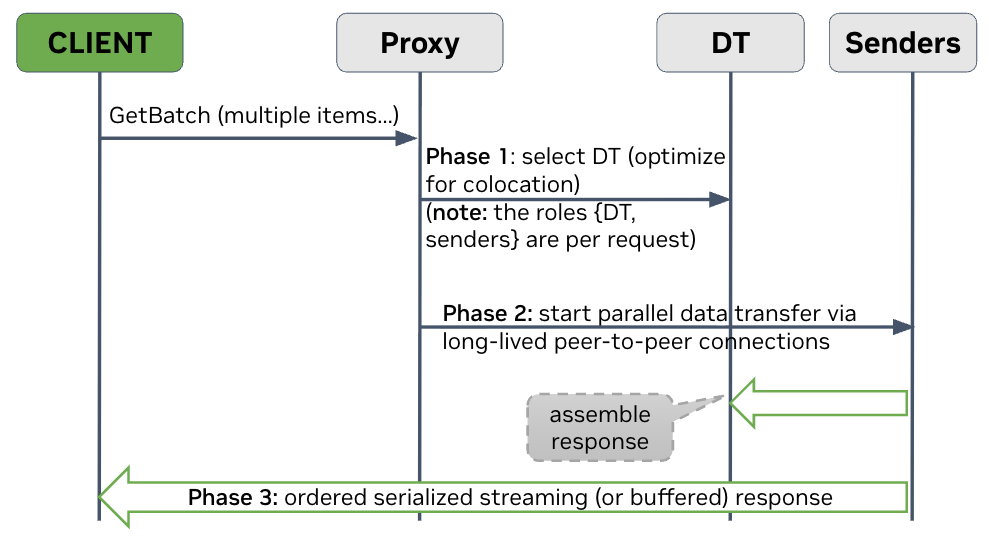}
  \caption{GetBatch execution model. A client submits a batch request to a proxy, which selects a Designated Target (DT). The proxy activates all other targets as senders. Senders stream locally owned data to the DT over peer-to-peer paths, and the DT emits a single output stream in strict request order.}
  \label{fig:getbatch:flow}
\end{figure}

\paragraph{Phase 1: DT Registration.}
The proxy forwards the request body to the selected DT. The DT allocates per-request execution state and returns a unique execution identifier, establishing itself as the sole coordinator responsible for producing the serialized output stream.

\paragraph{Phase 2: Distributed Sender Activation.}
After the DT accepts the request, the proxy broadcasts a control message to all other storage nodes - henceforth, \emph{senders}. Each sender independently determines which request entries it can satisfy locally—either by reading full objects it owns or by extracting specified members from locally stored archive shards—and begins delivering those payloads to the DT over persistent peer-to-peer connections.

Senders operate autonomously and in parallel. They do not coordinate with one another and may begin producing data as soon as local reads complete. Across the cluster, each storage node may simultaneously play multiple roles across concurrent requests (e.g., acting as DT for one GetBatch while serving as a sender for others), enabling high aggregate throughput under multi-tenant training workloads.

\paragraph{Phase 3: Client Redirection and Ordered Assembly.}
Once sender activation is complete, the proxy redirects the client to the DT. The DT assembles locally read and remotely received items, and serves the resulting serialized output stream strictly in client-specified order. While data may arrive at the DT out of order from multiple senders, output ordering is enforced unconditionally, decoupling heterogeneous read and transfer latencies from output determinism.

Depending on execution options, the DT may begin streaming to the client as soon as the first entries become available, overlapping retrieval, assembly, and consumption. This reduces latency to first byte and improves accelerator utilization while preserving deterministic semantics.

Data transfer between storage nodes relies on a shared pool of persistent peer-to-peer connections that are reused across requests and operations, with idle connections reclaimed after a configurable timeout. This amortizes connection setup cost and avoids connection storms under concurrent load.



\subsection{GetBatch Execution Options and Capabilities}
\label{sec:getbatch:options}

Beyond its core execution model, GetBatch exposes a small set of execution options and system capabilities that allow applications and operators to trade off latency, robustness, and data movement cost. These options do not affect correctness: regardless of configuration, GetBatch preserves strict ordering and deterministic output semantics. Instead, they control how the system executes under varying workload characteristics and failure conditions.
\subsubsection{Request-level Execution Options}

GetBatch requests may specify execution options that modify delivery behavior while preserving semantic guarantees.

\paragraph{Streaming (\texttt{strm}).}
When enabled, the DT begins emitting the output stream as soon as the earliest entries become available, overlapping retrieval, assembly, and consumption. This reduces time to first byte and improves accelerator utilization. When disabled, the DT buffers the entire result prior to delivery.

\paragraph{Continue-on-error (\texttt{coer}).}
By default, retrieval errors abort the request. When continue-on-error is enabled, recoverable per-entry failures (e.g., missing objects) do not terminate execution. Instead, failed entries are surfaced as explicit placeholders in the output stream, preserving positional correspondence with the request. Details of soft error classification and recovery are described in \Cref{sec:getbatch:faults}.

\paragraph{Colocation hints (\texttt{coloc}).}
By default, GetBatch routes requests opaquely: the proxy selects a DT without inspecting the request body, avoiding the cost of unmarshaling potentially large entry lists. When clients provide a colocation hint via a query parameter, the proxy unmarshals the request and computes per-entry placement weights to select the DT that owns the largest fraction of requested data, reducing cross-node transfers. This two-tier approach ensures that the common case pays no coordination overhead, while structured workloads can opt in to placement-aware routing when the reduction in network traffic justifies the additional proxy-side processing.

Conceptually, colocation captures two dimensions of containment common in machine learning datasets: (1) objects tend to be clustered on a subset of storage nodes, and (2) samples are often grouped within a small number of archive shards. Exploiting this structure can reduce network traffic and shard re-open costs.

\subsubsection{Fault Handling and Completion}
\label{sec:getbatch:faults}

Errors encountered during retrieval are classified on a per-entry basis and reported to the Designated Target (DT). GetBatch distinguishes between hard errors, which abort the request, and soft errors, which may be tolerated depending on request options. Soft errors include missing objects or archived files, transient failures of peer-to-peer data streams, and timeouts while waiting for remote senders.

When continue-on-error is enabled, soft errors do not terminate execution. Instead, the DT records the failure and emits a placeholder entry in the serialized output stream while preserving strict positional correspondence with the request. This allows downstream consumers to detect and handle missing samples without breaking batch alignment. Configurable soft error handling is intended to prevent premature termination of long-running training jobs, which may span many hours, due to a small number of missing or transiently unavailable samples.

To prevent unbounded degradation, GetBatch enforces configurable limits on recoverable failures and recovery attempts. Once these limits are exceeded, further failures are treated as fatal and the request is aborted. Upon successful completion or termination, the DT finalizes the serialized output stream and releases all per-request execution state.

\subsubsection{Configuration and Admission Control}

GetBatch exposes a dedicated configuration section that governs execution behavior under load. Configurable parameters include the maximum time the DT waits for a remote sender before initiating recovery, the number of get-from-neighbor (GFN) recovery attempts permitted per request, the maximum number of tolerated soft errors per request, and the number of background read-ahead workers used to warm the page cache for upcoming local reads.

Admission control is enforced at the DT to prevent resource exhaustion. Memory pressure is treated as a hard constraint: when memory utilization reaches a critical threshold, new work items are rejected with HTTP 429 (Too Many Requests), allowing clients to back off and retry. CPU and disk pressure are handled via throttling—the DT inserts calibrated sleep intervals to provide backpressure while allowing in-flight work to make forward progress. These controls allow operators to tune the balance between throughput, latency, and fault tolerance without modifying application logic.

\subsubsection{Observability and Monitoring}
\label{sec:getbatch:observability}

GetBatch exposes lightweight, per-node Prometheus metrics that characterize both \emph{workload composition} and \emph{execution bottlenecks}. At the request level, the system reports the total number of executed work items and the number and cumulative size of delivered items, separating whole-object retrieval from shard extraction. This distinction allows operators to quantify the fraction of work spent on shard extraction versus object delivery.

To diagnose performance bottlenecks, GetBatch separates time spent \emph{waiting for peer senders} from time spent \emph{throttling under resource pressure}. Specifically, it exports cumulative time waiting to receive entries from peer targets (\texttt{rxwait}) and cumulative time slept due to local pressure (\texttt{throttle}). This decomposition helps distinguish network- or skew-induced delays, such as slow or overloaded senders, from local bottlenecks at the Designated Target (DT), including CPU, disk, or memory pressure, guiding targeted remediation.

Finally, GetBatch reports error counters that distinguish \emph{hard failures}, including request failures and admission rejections, from \emph{soft errors} tolerated under configured limits, along with recovery activity such as total recovery attempts and failed recoveries. Together, these metrics support online monitoring of multi-tenant batch retrieval workloads and provide actionable signals for tuning timeouts, recovery bounds, and admission control.

\subsection{GetBatch Client-side Interface}
\label{sec:methods:getbatch_client_side}

From the client perspective, GetBatch exposes batch retrieval as a single logical operation. Rather than issuing individual object requests, the client constructs a batch request \emph{after sampling}, specifying the exact set of samples required for the current training step. GetBatch preserves a clean separation between \emph{sampling} and \emph{data access}: sampling logic---including shuffling, bucketing, and batch formation---remains entirely client-side, while the storage system handles retrieval.

AIStore provides client-side abstractions through its Python SDK\footnote{\url{https://pypi.org/project/aistore}} that integrate directly with data loaders:

\begin{lstlisting}[language=Python]
from aistore.sdk import Client

client = Client("http://ais-gateway")
bucket = client.bucket("training-data")

batch = client.batch(["obj_1", "obj_2", "...", "obj_n"], bucket)
for obj_info, content in batch.get():
    process_sample(content)
\end{lstlisting}

This pattern integrates into existing training frameworks with minimal changes: only the data access path is modified, while sampling and training logic remain unchanged. Detailed API documentation is provided in the AIStore GetBatch documentation.\footnote{\url{https://github.com/NVIDIA/aistore/blob/main/docs/get_batch.md}}

\section{Synthetic Benchmark}
\label{sec:synthetic}

We first evaluate GetBatch using a controlled synthetic benchmark.
This benchmark isolates the performance characteristics of batched retrieval across varying object sizes and batch configurations under fixed concurrency and hardware conditions.

\textbf{Cluster configuration:}
All experiments were conducted on a 16-node AIStore deployment hosted on Oracle Cloud Infrastructure (OCI). Each node runs one AIStore proxy and one target (16 proxies, 16 targets total) with the following resources:
\begin{itemize}[noitemsep, topsep=2pt]
    \item Instance type: BM.DenseIO.E5.128
    \item CPU: 128 OCPUs; Memory: 1536~GB
    \item Storage: 12 $\times$ 6.8~TB NVMe SSDs (81.6~TB per node)
    \item Network: 1 $\times$ 100~Gbps NIC
\end{itemize}
The total cluster capacity is 1.16~PiB across 192 NVMe drives.

\subsection{Experimental Setup}
\label{sec:synthetic:setup}

\paragraph{Benchmark Tool and Workload.}
We use AISLoader\footnote{\url{https://github.com/NVIDIA/aistore/blob/main/docs/aisloader.md}}, a load-generation tool designed for AIStore and S3-compatible object storage systems.
To avoid client-side bottlenecks, the workload is generated from \textbf{8} dedicated client nodes with an identical hardware configuration to the cluster nodes.
Each client node runs 10 concurrent workers, resulting in a total of 80 concurrent workers issuing retrieval requests.

\paragraph{Experimental procedure.}
For each object size (10~KiB, 100~KiB, 1~MiB), we execute:
\begin{enumerate}[noitemsep, topsep=2pt]
    \item Individual GET per object (baseline)
    \item GetBatch with batch size 32
    \item GetBatch with batch size 64
    \item GetBatch with batch size 128
\end{enumerate}
Each configuration runs for 1 hour to capture steady-state behavior.
Page caches are cleared on all cluster and client nodes before each run (to eliminate page-caching effects).

\begin{figure*}[t]
\centering
\includegraphics[width=0.9\textwidth]{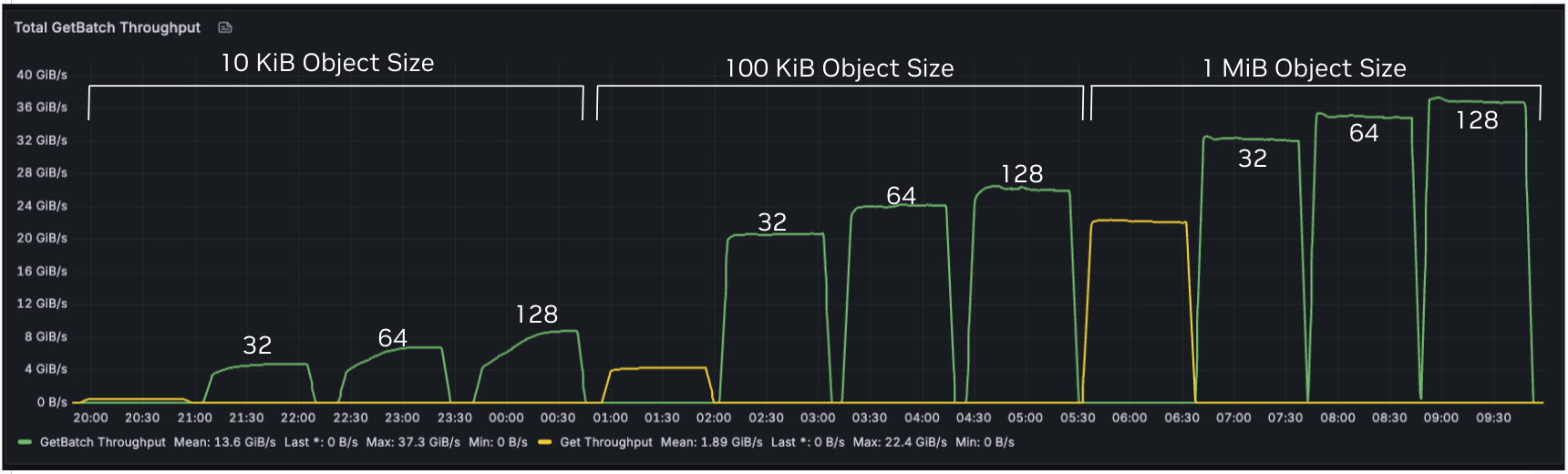}
\caption{Sustained throughput comparison between individual GET and GetBatch across object sizes and batch configurations. GetBatch yields the largest gains for small objects, where per-request overhead dominates.}
\label{fig:getbatch_throughput}
\end{figure*}

\subsection{Results}
\label{sec:synthetic:results}

\Cref{tab:getbatch_throughput} reports sustained throughput for individual GET and GetBatch across object sizes and batch configurations, while \Cref{fig:getbatch_throughput} visualizes the corresponding scaling trends.

\begin{table}[t]
\centering
\small
\setlength{\tabcolsep}{4pt}
\caption{Throughput (GiB/s) for individual GET and GetBatch. Speedup over GET in parentheses.}
\label{tab:getbatch_throughput}
\begin{tabular}{lcccc}
\toprule
 & & \multicolumn{3}{c}{\textbf{GetBatch}} \\
\cmidrule(lr){3-5}
\textbf{Object Size} & \textbf{GET} & \textbf{Batch 32} & \textbf{Batch 64} & \textbf{Batch 128} \\
\midrule
10 KiB  & 0.5 & 4.5 (9$\times$)  & 6.0 (12$\times$) & 7.3 (15$\times$) \\
100 KiB & 4.2 & 20.7 (4.9$\times$) & 24.1 (5.7$\times$) & 26.1 (6.2$\times$) \\
1 MiB   & 22.3 & 32.4 (1.5$\times$) & 35.2 (1.6$\times$) & 37.0 (1.7$\times$) \\
\bottomrule
\end{tabular}
\end{table}

GetBatch consistently outperforms individual GET across all evaluated object sizes.
The relative improvement decreases as object size increases, reflecting a transition from request-overhead dominance to data-transfer dominance.

\paragraph{10~KiB objects.}
GetBatch achieves up to 15$\times$ throughput improvement over individual GET.
At this size, per-request overheads---TCP round trips, request parsing, and scheduling---dominate total latency.
GetBatch amortizes these overheads across the batch.
\paragraph{100~KiB objects.}
GetBatch delivers a 6.2$\times$ improvement.
Data transfer time becomes more significant, but batching continues to reduce request-level overhead and improve network utilization.
\paragraph{1~MiB objects.}
GetBatch achieves a 1.7$\times$ improvement.
At this size, data transfer dominates total latency, reducing the relative benefit of batching.

\section{End-to-End Training: Canary-1B-Flash}
\label{sec:e2e}

The synthetic benchmark isolates GetBatch performance under controlled conditions with uniform object sizes and fixed concurrency. We next evaluate GetBatch in a production-scale training workload, where object sizes vary, access patterns are determined by the training sampler, and data loading must sustain GPU utilization under real scheduling constraints. All experiments in this section use the same 16-node AIStore cluster and hardware configuration described in \Cref{sec:synthetic}.

\subsection{Experimental Setup}
\label{sec:e2e:setup}

\textbf{Canary-1B-Flash} is an encoder-decoder speech recognition and translation model\footnote{\url{https://huggingface.co/nvidia/canary-1b-flash}} trained on 85k hours of speech across four languages (English, German, Spanish, French).
The model is trained using NVIDIA NeMo toolkit~\cite{kuchaiev2019nemo} and Lhotse for data loading~\cite{zelasko2021lhotse}.
We use 128 NVIDIA A100 80GB GPUs for distributed data parallel training with dynamic bucketing and OOMptimizer~\cite{zelasko2025emmett} (an OOM-aware batch size optimizer) to maximize throughput.
For a detailed training setup description, we refer the reader to~\cite{puvvada2024less} and~\cite{zelasko2025training}.

We compare three data access configurations under identical model architecture, dataset, hardware, and hyperparameters:
\begin{enumerate}[noitemsep, topsep=2pt]
    \item \textbf{Sequential I/O} (baseline): entire archive shards are retrieved and samples selected sequentially;
    \item \textbf{Random access I/O (GET)}: only sampled audio files are extracted from archives via individual GET requests\footnote{\url{https://aistore.nvidia.com/docs/archive}};
    \item \textbf{Batched random access I/O (GetBatch)}: all samples for a training batch are retrieved in a single GetBatch request.
\end{enumerate}

\subsection{Latency Analysis}
\label{sec:e2e:latency}

With 16 A100 nodes (128 GPUs, 1,024 data loader workers), all three configurations deliver similar aggregate throughput. At this scale, the storage cluster has ample spare capacity: network bandwidth, disk throughput, and CPU resources are all far from their limits, so the per-request overhead that GetBatch eliminates is not the bottleneck for overall bandwidth. However, the structural advantages of GetBatch---fewer requests, less network chatter, and lower scheduling overhead---appear clearly in request-level latency.

Training efficiency is governed not only by aggregate bandwidth but also by latency---particularly tail latency. High P95 and P99 batch latency directly translate to delayed training steps and reduced GPU utilization. Latency improvements are most significant when the storage cluster is not continuously saturated---that is, when I/O queues at individual storage nodes are not kept full at all times. In practice, this is common: synchronous training loops create bursty access patterns with idle periods between gradient steps, smaller-scale deployments may not generate sufficient concurrency to saturate storage, and even large training runs experience phases of reduced I/O activity during checkpointing or evaluation. Under sustained saturation, pipelining masks per-request latency and throughput becomes the dominant metric.

\subsubsection{Setup}

To evaluate latency under higher per-node contention, we use a reduced client configuration: 4 NVIDIA A100 nodes (32 GPUs, 256 data loader workers) against the same 16-node AIStore cluster. All other parameters---dataset layout, model configuration, storage cluster configuration, and software versions---remain identical. Only the data access method differs. We compare the same three data access strategies defined in \Cref{sec:e2e:setup}.

\paragraph{Latency measurement.}
Latency is measured as the total time from when the client issues a request until all requested bytes are received. This includes request transmission, server-side processing, data transfer, and complete reception at the client.

We report:
\begin{itemize}[noitemsep, topsep=2pt]
    \item \textbf{Batch latency:} Time to retrieve all samples required for a training batch.
    \item \textbf{Per-object latency:} Effective time per individual sample.
\end{itemize}

\subsubsection{Results}

\begin{table}[t]
\centering
\small
\setlength{\tabcolsep}{5pt}
\caption{Latency comparison during training. Values in milliseconds (ms).}
\label{tab:latency_comparison}
\begin{tabular}{lcccc}
\toprule
\textbf{Method} & \textbf{P50} & \textbf{P95} & \textbf{P99} & \textbf{Avg} \\
\midrule
\multicolumn{5}{c}{\textbf{Batch Latency (ms)}} \\
\midrule
Sequential I/O & 243.7 & 431.2 & 638.9 & 261.4 \\
Random GET     & 934.7 & 3668.7 & 4814.3 & 1320.0 \\
GetBatch       & 427.5 & 1808.6 & 2744.7 & 624.7 \\
\midrule
\multicolumn{5}{c}{\textbf{Per-Object Latency (ms)}} \\
\midrule
Sequential I/O & 1.2 & 5.2 & 6.8 & 2.0 \\
Random GET     & 9.1 & 27.3 & 53.5 & 12.3 \\
GetBatch       & 5.1 & 10.5 & 14.5 & 5.7 \\
\bottomrule
\end{tabular}
\end{table}

\paragraph{Sequential I/O.}
Sequential I/O achieves the lowest median batch latency (243.7\,ms) because it performs a single GET to fetch a shard and then reads samples sequentially from the open connection. This eliminates per-sample request overhead entirely. However, sampling flexibility is constrained: batches must draw samples from the retrieved shard, and sampling across shards requires additional shard downloads. Per-object latency reflects sequential read from an open stream rather than an independent retrieval, and is therefore not directly comparable to the per-object latency of Random GET or GetBatch.

\paragraph{Random Access (GET).}
Random GET exhibits markedly higher tail latency. Each batch requires hundreds of independent GET operations, and batch completion time is determined by the slowest individual request. At P95, batch latency reaches 3,668.7\,ms, and at P99, 4,814.3\,ms. Per-object tail latency also increases substantially (P99: 53.5\,ms), indicating exacerbated straggler effects due to request-level variability.

\paragraph{GetBatch.}
GetBatch substantially reduces both median and tail latency relative to Random GET. Batch-level P95 latency decreases from 3,668.7\,ms to 1,808.6\,ms (2.0$\times$ reduction), and P99 latency improves from 4,814.3\,ms to 2,744.7\,ms (1.75$\times$ reduction). Average batch latency decreases from 1,320.0\,ms to 624.7\,ms (2.1$\times$ reduction).

Per-object improvements are even more pronounced at the tail: P99 latency drops from 53.5\,ms to 14.5\,ms (3.7$\times$ reduction).

The structural difference is critical: instead of issuing hundreds of independent GET requests per batch, GetBatch performs a single coordinated retrieval and streams all results through a single response stream. This reduces request amplification, lowers control-plane overhead, minimizes network chatter, and mitigates straggler amplification.

\paragraph{Implications for training stability.}
In synchronous data-parallel training, all workers must complete their batch load before the next gradient step can begin; the slowest worker determines step time. Batch retrieval latency thus directly governs step-time variability. The absolute spread between P99 and P50 batch latency quantifies this variability: for Random GET, the spread is 3,880\,ms (4,814.3 $-$ 934.7); for GetBatch, it narrows to 2,317\,ms (2,744.7 $-$ 427.5)---a 40\% reduction. In practical terms, the worst-case stall time drops from nearly 5 seconds to under 3 seconds. This tighter latency distribution reduces step-time jitter and improves GPU utilization regardless of whether the storage cluster is saturated: fat tails in batch retrieval propagate to idle GPU cycles, so reducing them improves training efficiency at any scale.

\section{Discussion}
\label{sec:discussion}

\subsection{When Does GetBatch Help?}
\label{sec:discussion:when}

GetBatch helps most when per-request overhead dominates data transfer time. For small objects (10--100~KiB), overhead from TCP round trips, request parsing, and per-request scheduling accounts for most of the retrieval time; batching amortizes this across the entire batch. As object size grows, data transfer dominates and the relative benefit diminishes---consistent with the observed decline from 15$\times$ at 10~KiB to 1.7$\times$ at 1~MiB in the synthetic benchmark (\Cref{sec:synthetic:results}). Even when aggregate throughput is comparable, as in the Canary training experiment (\Cref{sec:e2e:latency}), GetBatch delivers measurable latency improvements: a 2$\times$ reduction in P95 batch latency and a 3.7$\times$ reduction in P99 per-object latency (\Cref{sec:e2e:latency}). This profile aligns well with typical ML training datasets containing images, audio segments, and text samples.

\subsection{Scalability Considerations}
\label{sec:discussion:scalability}

Each GetBatch request is coordinated by a single Designated Target (DT)---a storage node (randomly) selected on a per-request basis to assemble the serialized output stream. The DT serves as the serialization point, which raises questions about scalability. In practice, three factors mitigate this concern. First, the DT performs only ordering and serialization; the compute-intensive work of reading data from storage is distributed across all senders. Second, different requests are routed to different DTs via consistent hashing, distributing the serialization load across the cluster. Third, streaming mode allows the DT to begin emitting output before all items arrive, reducing memory pressure and enabling pipelining.

Nevertheless, at very high concurrency or with extremely large batches, the DT could become a bottleneck. Under sustained synthetic load (\Cref{sec:synthetic}), we observe that disk utilization saturates first as the DT serves both local reads and incoming sender streams, followed by elevated memory and CPU pressure from buffering out-of-order arrivals and enforcing output order. Once these thresholds are reached, the admission control and throttling mechanisms described in \Cref{sec:getbatch:options} engage, applying backpressure to prevent resource exhaustion while allowing in-flight requests to complete. This degradation is graceful, but it bounds the throughput a single DT can sustain. Further investigation of DT scaling behavior at larger cluster sizes (32+ nodes) and higher concurrency levels is an important direction for future work.

\subsection{Client-side Integration}
\label{sec:discussion:integration}

Beyond throughput, GetBatch simplifies client-side data loading code. Without batched retrieval, a data loader must manage concurrent connections, handle per-object errors, and reassemble results in the correct order. With GetBatch, the loader submits a single batch specification and iterates over an ordered stream.

AIStore provides client-side abstractions through its Python SDK\footnote{\url{https://pypi.org/project/aistore}} that integrate directly with data loaders:

\begin{lstlisting}[language=Python,caption={Python pseudo-code illustrating the decoupling of data sampling and fetching using GetBatch. The iterable dataset samples and collects an entire batch in one step.},label={lst:dataloader}]
from torch.utils.data import (
  IterableDataset,
  DataLoader
)

class MyDataset(IterableDataset):
  def __next__(self):
    # State was initialized in __iter__()
    selected_paths = self._index.sample(n=self.batch_size)
    batch = client.batch(
      objects=selected_paths, 
      bucket=bucket,
    )
    tensors = []
    for metadata, content in batch.get():
      tensors.append(
        self._raw_data_to_tensor(content)
      )
    return torch.stack(tensors)

dl = DataLoader(
  MyDataset(...), batch_size=None,
)
      
\end{lstlisting}

The pattern in Listing~\ref{lst:dataloader} has been adopted in the Lhotse speech data library~\cite{zelasko2021lhotse}, where a GetBatch wrapper called \texttt{AISBatchLoader} is called once to replace per-sample retrieval for an entire batch. The loader collects all object references from the batch manifest, issues one GetBatch request, and injects the returned content directly into in-memory data structures---eliminating per-sample I/O management from training code entirely.

\subsection{Comparison with Alternative Approaches}
\label{sec:discussion:alternatives}

Several alternative mechanisms could reduce per-request overhead without a storage-native batching primitive. \textbf{HTTP/2 multiplexing} allows multiple requests over a single TCP connection, reducing connection setup overhead but not per-request parsing or server-side scheduling. \textbf{gRPC streaming} provides efficient bidirectional streaming but still requires per-object request-response cycles on the server side. \textbf{Client-side caching} avoids repeated fetches but does not reduce first-access latency and increases client memory requirements. These mechanisms address subsets of the overhead that GetBatch eliminates end-to-end. By performing server-side assembly, the storage system coordinates retrieval across nodes and delivers a single ordered response, amortizing not just connection overhead but also request parsing, scheduling, and intra-cluster data movement.

\subsection{Limitations and Tradeoffs}
\label{sec:discussion:limitations}

GetBatch involves tradeoffs. The current implementation requires AIStore as the storage backend, as batched retrieval semantics are not part of standard S3-compatible APIs. More broadly, the lack of server-side batch retrieval in modern object storage systems and data loading frameworks represents a gap in the current state of the art. We believe this capability would benefit the broader storage ecosystem and hope that other open-source projects and commercial vendors will adopt similar primitives. The current design enforces strict output ordering to support reproducible training; server-side shuffle modes are a natural extension that could further reduce client-side complexity for workloads that do not require deterministic sample order.

\section{Related Work}
\label{sec:related_work}

\subsection{Small-Scale Data Loading with Map-Style Datasets}
\label{sec:related_work:map_style}

Traditional deep learning workflows commonly employ PyTorch's map-style datasets, which implement the \texttt{\_\_getitem\_\_()} and \texttt{\_\_len\_\_()} protocols to provide random access to individual samples~\cite{paszke2019pytorch}. In this paradigm, datasets maintain an index-to-sample mapping where \texttt{dataset[idx]} retrieves the \texttt{idx}-th element by reading its file path from disk. This approach works well for small to medium-scale datasets that fit within available storage and where random access patterns do not significantly impact I/O performance. The standard PyTorch DataLoader consumes map-style datasets by iterating through a sampler that produces indices, effectively yielding \texttt{collate\_fn([dataset[i] for i in indexes])} for each batch. However, this index-based random access pattern becomes a significant bottleneck when scaling to large datasets, as random I/O operations are substantially slower and less efficient than sequential access patterns—often by a factor of 10-20x on traditional storage systems~\cite{leclerc2023ffcv,mohan2021analyzing}.

\subsection{Large-Scale Data Formats and Sequential I/O Optimization}
\label{sec:related_work:sequential_style}

To address the limitations of random access at scale, the machine learning community has developed several specialized data formats optimized for sequential I/O patterns. WebDataset stores samples in POSIX TAR archives, enabling efficient sequential reads that greatly speed up I/O operations on both rotational storage and networked file systems~\cite{aizman2019aistore}. By organizing data into shards (collections of samples stored together) WebDataset achieves parallel I/O while maintaining sequential access within each shard. Similarly, TensorFlow's TFRecord format stores sequences of binary protocol buffer records that can only be read sequentially, with sharding strategies enabling parallel data access and prefetching to reduce training step latency~\cite{abadi2016tensorflow}. FFCV introduces the .beton format, which stores data in a quasi-random order optimized for both sequential disk access and distributed training, achieving order-of-magnitude speedups on standard vision benchmarks~\cite{leclerc2023ffcv}. Parquet, a columnar storage format originally designed for data analytics, has been adapted for deep learning through libraries like Petastorm~\cite{yermolovich2018petastorm}, which enables efficient loading from Parquet files by reading only the necessary columns and leveraging the format’s metadata for selective access.

These formats fundamentally diverge from the assumptions of naive map-style dataloaders~\cite{paszke2019pytorch,hira2025spdl}. Because they prioritize sequential access patterns and are typically accessed through iterable-style datasets rather than random-access interfaces, standard index-based sampling strategies become incompatible. Iterable-style datasets provide an iterator over samples rather than supporting random access via \texttt{\_\_getitem\_\_()}, making them ideal for streaming large datasets that exceed memory capacity. 
Recent work has shown that data loading can become the primary bottleneck in modern training pipelines, with I/O stalls accounting for significant fractions of total training time~\cite{mohan2021analyzing,pumma2019lmdbio}. This shift from random to sequential access introduces new challenges for data shuffling and sampling, often requiring approximate shuffling through shuffle buffers or block-based sampling strategies that balance I/O efficiency with training randomness~\cite{leclerc2023ffcv,hira2025spdl}.

\subsection{GPU-Accelerated Data Loading}

NVIDIA DALI~\cite{nvidia2018dali} and SPDL~\cite{hira2025spdl} accelerate data loading through GPU-based preprocessing and optimized CPU--GPU data transfer. These systems primarily target the transformation and transfer stages of the input pipeline. GetBatch addresses a different layer of the stack: it optimizes the storage retrieval stage by pushing batching into the storage layer, reducing the number of network requests issued by the data loading framework.

Because these approaches operate at different stages of the pipeline, they are conceptually complementary. Exploring tighter integration between GetBatch and GPU-accelerated preprocessing frameworks such as DALI or SPDL is a promising direction for future work, potentially enabling end-to-end optimization from storage retrieval through transformation and accelerator transfer.

\section{Conclusions}
\label{sec:conclusions}

We introduced GetBatch, a storage primitive that treats batch retrieval as a first-class operation for machine learning data loading. By assembling all items required for a training batch into a single ordered response, GetBatch amortizes per-request overhead while preserving deterministic output ordering essential for reproducible training.

Our evaluation on a 16-node AIStore cluster demonstrates that GetBatch achieves up to 15$\times$ throughput improvement for small objects (10~KiB), where per-request overhead dominates, with consistent gains across batch sizes. For larger objects, GetBatch remains competitive with individual GET as data transfer cost becomes the primary factor. 

Beyond aggregate bandwidth, we show that GetBatch significantly improves latency stability during training. In a realistic distributed training setup, GetBatch reduces batch-level P95 latency by 2.0$\times$ and P99 latency by 1.75$\times$ compared to individual GET requests. Per-object tail latency improves by up to 3.7$\times$. By reducing request amplification and collapsing many independent object reads into a single coordinated retrieval, GetBatch lowers control-plane overhead, mitigates straggler effects, and improves training step-time stability.

GetBatch is open source and available as part of NVIDIA AIStore.\footnote{\url{https://github.com/NVIDIA/aistore}} The AISLoader benchmark tool used in the synthetic evaluation is included in the AIStore distribution.\footnote{\url{https://github.com/NVIDIA/aistore/blob/main/docs/aisloader.md}} GetBatch documentation, including API reference and integration guides, is available online.\footnote{\url{https://github.com/NVIDIA/aistore/blob/main/docs/get_batch.md}}

\section*{Acknowledgements}

We thank the AIStore engineering team for infrastructure support, and the NVIDIA NeMo, ASR Speech, and Lhotse teams for their collaboration and assistance with training integration.

\bibliography{example_paper}
\bibliographystyle{icml2026}

\end{document}